\documentclass[twocolumn]{NobArticle}

\usepackage[utf8]{inputenc}
\usepackage[ruled,vlined]{algorithm2e}


\runninghead{3D Blocking for Matrix-free Smoothers in 2D Variable-Viscosity Stokes Equations With Applications to Geophysics}

\title{3D Blocking for Matrix-free Smoothers in 2D Variable-Viscosity Stokes Equations with Applications to Geodynamics}

\author{
    Marcel Ferrari\textsuperscript{1}, 
    Cyrill Püntener\textsuperscript{2}*, 
    Alexander Sotoudeh\textsuperscript{2}*
    and Niklas Viebig\textsuperscript{3}*
}

\date{
    \textsuperscript{1} Department of Mathematics (D-MATH), ETH Zurich \\ 
    \textsuperscript{2} Department of Computer Science (D-INFK), ETH Zurich \\ 
    \textsuperscript{3} Department of Physics (D-PHYS), ETH Zurich \\ 
    * These authors contributed equally and are listed alphabetically by last name.
}
\renewcommand{\maketitlehookd}{%
\begin{abstract}
\noindent We present the design, implementation, and evaluation of optimized matrix-free stencil kernels for multigrid smoothing in the incompressible Stokes equations with variable viscosity, motivated by geophysical flow problems. We investigate five smoother variants derived from different optimisation strategies: Red–Black Gauss–Seidel, Jacobi, fused Jacobi, blocked fused Jacobi, and a novel Jacobi smoother with RAS-type temporal blocking, a strategy that applies local iterations on overlapping tiles to improve cache reuse. To ensure correctness, we introduce an energy-based residual norm that balances velocity and pressure contributions, and validate all implementations using a high-contrast sinker benchmark representative of realistic geodynamic numerical models. Our performance study on NVIDIA GH200 Grace Hopper nodes of the ALPS supercomputer demonstrates that all smoothers scale well within a single NUMA domain, but the RAS-Jacobi smoother consistently achieves the best performance at higher core counts. It sustains over $90\%$ weak-scaling efficiency up to 64 cores and delivers up to a threefold speedup compared to the C++ Jacobi baseline, owing to improved cache reuse and reduced memory traffic. These results show that temporal blocking, already employed in distributed-memory solvers to reduce communication, can also provide substantial benefits at the socket and NUMA level. This work highlights the importance of cache-aware stencil design for harnessing modern heterogeneous architectures and lays the groundwork for extending RAS-type temporal blocking strategies to three-dimensional problems and GPU accelerators.
\medskip
\end{abstract}
}

\begin{document}

\small
\maketitle

\section{Introduction}
The Stokes equations are widely used in various scientific and engineering applications to describe the flow of slow-moving, highly viscous fluids. These equations result from a linearization of the full Navier–Stokes equations, under the assumption that inertial forces are negligible in comparison to viscous forces~\cite{Gerya_stokes_flow}. In the field of geophysics, the Stokes equations play a crucial role, particularly under the Boussinesq approximation, where they are used to model complex geodynamic phenomena such as the dynamics of subduction zones~\cite{Gerya_boussinesq}. Over the years, numerous numerical models have been developed to simulate thermomechanically coupled systems that deform according to the Stokes flow equations. A wide range of discretization techniques has been employed for this purpose, including finite elements~\cite{ptatin3d} and finite difference methods~\cite{gerya_thermomechanical, TACKLEY20087}. Among these, a particularly successful family of models has emerged from the work of Taras Gerya over the past two decades. These models are based on a marker-in-cell approach (also commonly known as the material-point method), which combines particle-based material advection with a finite difference discretization of the governing differential equations~\cite{gerya_thermomechanical}. This modeling framework has been extended to capture a variety of complex physical processes, such as including a fully visco-elasto-plastic rheology model~\cite{Gerya_viscoelastoplastic}, two-phase hydro-mechanical flow~\cite{zilio_hydro_mechanical}, and the evolution of material grain size~\cite{gerya_nature}, among others. Many of these models rely on the assumption of incompressibility, or at most weak compressibility of the material. As a result, the discretization of the Stokes flow equations typically leads to saddle-point linear systems, which is well-known for being particularly challenging and expensive to solve numerically~\cite{saddle_point}.

Given the widespread success of these numerical models, there is an ongoing effort to modernize the existing codebase by porting it to the Python programming language. This initiative is driven by several key motivations, including the desire to increase productivity and simplicity by leveraging Python’s extensive scientific ecosystem. Additionally, it aims to improve performance through enhanced GPU support and to streamline the integration of machine learning models for capturing unknown or poorly understood physical processes through data-driven approaches. The new codebase, named Pyroclast~\cite{ferrari2025geodynamic}, seeks to implement the core functionalities of this family of models while providing researchers with a flexible and efficient platform to prototype and test new ideas rapidly.

\subsection{Contributions}
\label{sec:contributions}
This work makes the following contributions:
\begin{itemize}
    \item We design and implement a family of matrix-free stencil kernels for velocity smoothing in multigrid solvers for the incompressible Stokes equations with variable viscosity.
    \item We introduce a RAS-type temporal blocking strategy, adapted from domain decomposition methods, that improves cache reuse and scalability within a single node.
    \item We implement and evaluate five distinct smoother variants (RBGS, Jacobi, Fused Jacobi, Blocked Fused Jacobi, and RAS-Jacobi) in both Python (Numba) and C++.
    \item We propose an energy-based residual norm that balances velocity and pressure components, enabling fair convergence assessment in problems with large viscosity contrasts.
    \item We validate correctness using a high-contrast sinker benchmark and demonstrate stable convergence across all smoother variants.
    \item We perform strong and weak scaling experiments on NVIDIA GH200 Grace Hopper nodes of the ALPS supercomputer, showing that the RAS-Jacobi smoother achieves the best performance, sustaining over 90\% weak-scaling efficiency up to 64 cores and delivering up to $3\times$ speedup over the classic Jacobi baseline.
\end{itemize}

\subsection{Paper Organisation}
The remainder of this paper is organised as follows.  
Section~\ref{sec:numerical_model} introduces the governing equations and the numerical model implemented in the Pyroclast framework.  
Section~\ref{sec:fast_solver} presents the iterative strategy used to solve the Stokes system, including Uzawa iteration, inexact Uzawa, and Anderson acceleration.  
Section~\ref{sec:methods} discusses the optimisation strategies investigated for our stencil kernels, while Section~\ref{sec:algorithms} describes the smoother implementations derived from these strategies.  
Section~\ref{sec:correctness} validates the correctness of the methods through the high-contrast sinker benchmark, including solver configuration, smoother setup, residual measurement, and results.  
Section~\ref{sec:benchmarks} reports the synthetic benchmarks, detailing the experimental setup, system configuration, and both strong and weak scaling results.  
Finally, Section~\ref{sec:conclusions} concludes the paper and Section~\ref{sec:future_work} outlines directions for future research.

\section{Numerical Model}
\label{sec:numerical_model}

Pyroclast implements a variety of numerical models in an object-oriented framework, allowing for flexibility when integrating new physics. At its core, the base mechanical model solve a buoyancy-driven flow problem using the marker-in-cell method in order to capturing highly complex material behavior and avoid numerical diffusion~\cite{Gerya_MIC}. This section focuses on describing the specific numerical model employed to solve buoyancy-driven flows under the extreme physical conditions typically encountered in geodynamic systems.

\subsection{The Incompressible Stokes Equations}
The two-dimensional incompressible Stokes equations consist of a coupled system of three partial differential equations: the x-momentum equation, the y-momentum equation, and the incompressibility condition. In geophysical applications, it is essential to employ a stress-conservative discretization of the momentum equations to ensure physically accurate results, particularly when modeling complex material behavior. Consequently, we adopt the Cauchy formulation of the Stokes equations, expressing the problem in terms of deviatoric stresses and pressure as follows:

\begin{gather}
\frac{\partial \sigma_{xx}'}{\partial x} + \frac{\partial \sigma_{xy}'}{\partial y} - \frac{\partial p}{\partial x} = 0 \label{eq:xmom} \\
\frac{\partial \sigma_{yx}'}{\partial x} + \frac{\partial \sigma_{yy}'}{\partial y} - \frac{\partial p}{\partial y} = \rho g_y \label{eq:ymom} \\
\sigma_{ij}' = 2 \eta \dot \epsilon_{ij} = \eta \left(\frac{\partial v_i}{\partial x_j} +  \frac{\partial v_j}{\partial x_i}\right) \label{eq:stress} \\
\frac{\partial v_x}{\partial x} + \frac{\partial v_y}{\partial y} = 0 \label{eq:incomp}
\end{gather}

Equations~\eqref{eq:xmom} and~\eqref{eq:ymom} represent the x- and y-momentum balances, respectively. Equation~\eqref{eq:stress} defines the constitutive relationship between deviatoric stress and strain rate under the assumption of a Newtonian viscous rheology, and Equation~\eqref{eq:incomp} enforces the incompressibility condition.
The physical quantities in the above equations are defined as follows:

\begin{itemize}
\item $\sigma_{ij}'$: deviatoric stress on the $i$-plane acting in the $j$-direction
\item $p$: pressure
\item $v_x, v_y$: velocity components in the $x$ and $y$ directions, respectively
\item $\rho$: density
\item $g_y$: gravitational acceleration in the $y$-direction
\item $\eta$: viscosity
\item $\dot \epsilon_{ij}$: strain-rate on the $i$-plane acting in the $j$-direction
\end{itemize}

\begin{figure}
    \centering
    \includegraphics[width=0.95\linewidth]{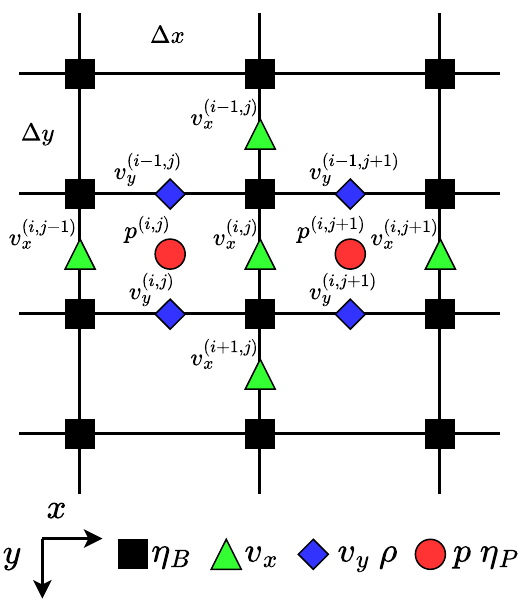}
    \caption{Staggered grid stencil for the discretization of the $x$-momentum equation. Velocity components $v_x$ (green triangles) and $v_y$ (blue diamonds) are located at vertical and horizontal cell faces, respectively, while pressure $p$ (red circles) is stored at cell centers. Viscosity is defined at basic nodal points $\eta_B$ (black squares) and at pressure points $\eta_P$, with density $\rho$ collocated with $v_y$. Nodes carrying multiple labels indicate co-located quantities following the indexing scheme of the primary solution variables. This illustration is adapted from~\cite{Gerya_stokes_flow}.
}
    \label{fig:x-mom-stencil}
\end{figure}

The discretization of the Stokes equations is performed using finite differences on a fully staggered grid. This grid arrangement ensures that all partial derivatives are approximated consistently using central difference schemes, which improves the accuracy and stability of the numerical solution. \Cref{fig:x-mom-stencil} illustrates the stencil employed for the discretization of the x-momentum equation. It is important to note that both viscosity and density are treated as spatially variable parameters and are stored explicitly on the grid to capture variation of material properties.

Following the discretization process, the system of partial differential equations is transformed into a sparse linear system expressed in terms of the velocity and pressure unknowns.

\subsection{The Advection Equation}
The advection of material properties between time steps is performed from a Lagrangian perspective using a set of particles, commonly referred to as markers. The time-integration scheme follows an Eulerian-Lagrangian splitting approach, which decouples the advection of material properties from the solution of the mechanical problem. Material properties are initially prescribed on the markers, after which the time loop begins. At each time step, marker properties are first interpolated onto the Eulerian grid. The Stokes equations are then solved on this grid to obtain the velocity and pressure fields ($v_x$, $v_y$, and $p$). A suitable time step $\Delta t$ is determined based on physical criteria, after which the velocity field is interpolated back onto the markers. The markers are then advected using Heun’s explicit second-order Runge-Kutta method (RK2). Because this splitting method separates advection from the solution of the Stokes equations, the solver itself can operate in a ``black-box'' fashion, without any knowledge of the underlying markers during the solution phase.

\section{Fast Stokes Solver}
\label{sec:fast_solver}
The solution of the Stokes equations represents the dominant computational cost throughout the simulation. The discretization process results in a sparse linear system that is a saddle-point problem, a structure commonly encountered in systems that satisfy the optimality conditions of equality-constrained optimization problems. In this context, the pressure acts as a Lagrange multiplier, enforcing the incompressibility constraint on the velocity field~\cite{saddle_point}. In practice, the key challenge consists in determining values for pressure from this constraint, as the incompressibility condition contains no explicit pressure term. As a result, the discretized linear system features a pressure block with zero diagonal entries. This saddle-point structure leads to difficulties in both stability and efficiency of iterative solvers and requires specialized algorithms to ensure convergence. The next few sections will discuss the numerical methods implemented in Pyroclast for the solution of the Stokes system.
\subsection{Uzawa Iteration}
A common strategy for solving saddle-point systems such as those arising from the Stokes equations is to decouple the computation of velocity and pressure through a Schur complement reduction. This method is known as \textit{Uzawa iteration}.

The discrete Stokes system can be written in block form as a saddle-point system:
\begin{equation}
\label{eq:stokes_system}
\begin{bmatrix}
M & G \\
D & 0
\end{bmatrix}
\begin{bmatrix}
v \\
p
\end{bmatrix}
=
\begin{bmatrix}
f \\
0
\end{bmatrix},
\end{equation}
where $M$ represents the discrete momentum operator, $G$ is the discrete gradient operator, and $D$ is the discrete divergence operator.
The first row of this system yields the momentum balance equation:
\begin{equation}
\label{eq:velocity_subsystem}
M v + G p = f.
\end{equation}
Solving for $v$ explicitly gives:
\begin{equation} \label{eq:velocity_elimination}
v = M^{-1} (f - G p).
\end{equation}
Substituting equation~\eqref{eq:velocity_elimination} into the incompressibility constraint $D v = 0$ and rearranging yields the \textit{Schur complement system} for pressure:
\begin{equation}
\label{eq:schur_system}
D M^{-1} G p = D M^{-1} f,
\end{equation}
where $S = D M^{-1} G$ is the Schur complement. Assembling $S$ is an intractable problem, so in order to solve~\eqref{eq:schur_system} a Richardson fixed-point iteration scheme is used:
\begin{equation}
\label{eq:full_richardson_iter}
p^{k+1} = p^k + \alpha \left( D M^{-1} (f - G p^k) \right),
\end{equation}
where $\alpha$ is a relaxation parameter. In order to make this procedure computationally feasible, we use equation~\eqref{eq:velocity_elimination} and define:
\begin{equation}
v^{k+1} := M^{-1} (f - G p^k).
\end{equation}
Using this definition, we apply a preconditioner \mbox{$S^{-1} \approx  \eta $}, where $\eta $ is a diagonal matrix that approximates the inverse Schur complement via the viscosity field. Under this approximation, the pressure update in~\eqref{eq:full_richardson_iter} finally simplifies to:
\begin{equation}
p^{k+1} = p^k + \alpha \eta D v^{k+1}.
\end{equation}
This splitting defines the Uzawa scheme: at each iteration, the velocity is first updated by solving the momentum equation with the current pressure iterate, and the pressure is then corrected using the residual of the incompressibility condition scaled by the viscosity field.

\subsection{Inexact Uzawa Iteration}
In practice, computing an exact solution for the velocity update $v^{k+1}$ by performing a sparse factorization of the momentum matrix $M$ is prohibitively expensive. Instead, we seek an approximate solution for $v^{k+1}$. Given the simple rectangular grid used in our discretization, we can efficiently compute this approximate solution using a \textit{geometric multigrid} method~\cite{Gerya_multigrid}. Since the matrix $M$ is not a saddle-point matrix, classical smoothers such as Gauss-Seidel can be safely used within the multigrid hierarchy. Furthermore, we exploit the structured geometry of the grid to implement the multigrid solver in a fully matrix-free way by leveraging stencil operations. This approach of solving the velocity update approximately, rather than exactly, leads to what is known as the \textit{inexact Uzawa iteration}. In this context, the bulk of the computational cost is typically the smoothing iterations, which are applied repeatedly across the levels of the multigrid hierarchy and dominate the overall runtime of the Stokes solver. Consequently, it is crucial to implement these operations using highly optimized matrix-free stencil routines to achieve both computational efficiency and scalability on modern hardware architectures.
\subsection{Anderson Acceleration}
Although inexact Uzawa iteration provides an efficient way of decoupling the velocity and pressure solves, it still requires a sufficiently accurate solution of the momentum subsystem to ensure fast and reliable convergence. To accelerate this process, we apply Anderson acceleration to the fixed-point iteration defined by the inexact Uzawa scheme on the combined velocity-pressure space $x = [v, p]$. The mapping $F(x)$ represents one inexact Uzawa iteration, consisting of an approximate velocity solve followed by a pressure update. Anderson acceleration constructs the next iterate as a linear combination of previous iterates by minimizing the residual over the past $m$ iterations. 

\begin{algorithm}[!b]
\caption{Anderson Acceleration for Inexact Uzawa Iteration}
\label{alg:anderson}
\SetKwInOut{Input}{Input}
\SetKwInOut{Output}{Output}

\Input{Initial guess $x_0 = [v_0, p_0]$}
\Output{Approximate solution $x = [v, p]$}

Compute $x_1 := F(x_0)$ \tcp*{One Uzawa iteration}

\For{$k := 1, 2, \ldots$ until convergence}{
    $m_k := \min(m, k)$\;
    $r_k := F(x_k) - x_k$\;
    Form $R_k := [r_{k - m_k}, \ldots, r_k]$\;
    Find $\alpha_k := \arg \min_{\alpha \in A_k} \| R_k \alpha \|_2$\;
    where $A_k := \left\{\alpha \in \mathbb{R}^{m_k + 1} \, : \, \sum_{i=0}^{m_k} \alpha_i = 1 \right\}$\;
    $x_{k+1} := \sum_{i=0}^{m_k} (\alpha_k)_i F(x_{k - m_k + i})$\;
}
\end{algorithm}

Specifically, at each iteration, we collect the differences between the current iterate and the fixed-point mapping, \mbox{$r_k = F(x_k) - x_k$}, into a residual matrix $R_k$. The new iterate is then formed by solving a small least-squares problem to minimize the norm of the combined residual, subject to the constraint that the combination coefficients sum to one. This approach is outlined in Algorithm~\ref{alg:anderson}, where $m$ controls the depth of the history used for acceleration. This method significantly reduces the number of velocity solves required per Uzawa iteration, usually to a maximum of 2-3 cycles. It can be shown that this strategy results in a solver that is closely related to (flexible) GMRES with a matrix-splitting preconditioner~\cite{ho2016acceleratinguzawaalgorithm}.

\section{Optimisation Strategies}
\label{sec:methods}
Stencil kernels are central to the computational performance of our solver, as they are the dominant cost in the multigrid hierarchy. In order to exploit modern memory hierarchies and increase arithmetic intensity, we investigated several optimisation strategies. These techniques aim to minimise redundant memory traffic, improve cache locality, and increase computational throughput without sacrificing the flexibility of our implementation. The following subsections describe the three main optimisation strategies we explored. An in-depth discussion of possible optimization strategies is available in~\cite{stencil_optimizations}.

\subsection{Loop Fusion}
Our stencil kernels operate on the staggered velocity components $v_x$ and $v_y$. In a naive approach, the two updates are performed in separate sweeps over the computational grid, resulting in a full pass for $v_x$ followed by a full pass for $v_y$. By leveraging geometric information about the staggered layout, we fuse these operations into a single kernel that updates both velocity components simultaneously. This strategy reduces the number of memory loads required for the viscosity field, since the same values can be reused for both updates. The trade-off, however, is that the fused kernel increases in complexity, which may reduce the ability of the compiler to apply low-level optimisations such as vectorisation or instruction reordering.

\subsection{Spatial Blocking}
Since our solver frequently operates on large arrays that exceed cache capacity, memory locality becomes a limiting factor. To mitigate this, we introduced spatial blocking, partitioning the domain into subregions that can be processed independently. By reusing data that remains in cache across stencil operations within a block, this technique can reduce memory bandwidth pressure and improve overall performance. The effectiveness of spatial blocking depends on the block size relative to the cache hierarchy, as well as the specific access patterns of the staggered stencil.

\subsection{RAS-Type Temporal Blocking}
\label{sec:ras-type_temporal_blocking}
While two-dimensional blocking is often challenging to implement efficiently for complex stencils, temporal blocking can provide additional reuse of loaded data across iterations. We designed a three-dimensional blocking strategy that combines spatial and temporal blocking, inspired by the restricted additive Schwarz (RAS) preconditioning method. Classical temporal blocking schemes are mathematically exact but typically introduce significant implementation complexity, additional branching logic, and overhead that can hinder performance in kernels as intricate as ours (cfr.~\cite{stencil_optimizations}). In contrast, our approach is deliberately simple, intuitive, and easy to optimise (see Algorithm~\ref{alg:ras_temporal}).

Each tile is solved independently by imposing Dirichlet boundary conditions on its outer boundary, which lies beyond the overlap. Only the interior (non-overlapping) values are retained, while the overlap values are discarded. Any choice of local solver may be used, whether direct or iterative. A key feature of RAS is that the overlap values are computed redundantly and later discarded. This avoids directly interfacing active domain values with fixed and stale boundary conditions, instead providing flexibility through proxy values that are solved but not retained. In distributed-memory settings with very large subdomains, the cost of these redundant computations is negligible and the method effectively reduces communication overhead. However, in our case we operate at the level of cache-optimal tiles (e.g., $32 \times 32$ or $64 \times 64$), where halo values represent a significant fraction of the total work and thus introduce nontrivial computational overhead.

To mitigate this effect, we employ a strategy based on periodically and randomly shifted blocking. At each outer iteration, the tiling grid is shifted by a random offset in both spatial directions, with periodic wrapping. This ensures that the seams between tiles change dynamically, preventing the accumulation of artefacts caused by persistently stale values along fixed tile boundaries. Conceptually, iterating over tiles and applying our stencil-based iterative solver as the local solver corresponds to performing three-dimensional blocking on a two-dimensional domain, with the temporal dimension emerging from the local iterations.
The tiling logic follows a simple boundary-ownership convention in which each tile owns its top and left boundaries, while the bottom and right boundaries are owned by adjacent tiles. Because the stencil requires values beyond the tile interior, this leads to overlapping regions where data may be read concurrently. In particular, a tile may read from the bottom or right boundaries while those entries are being updated by a neighbouring tile. This introduces a race condition, but it is benign: each cell is written by exactly one tile, and concurrent reads of in-progress updates do not affect the stability or convergence of the method.
This perspective provides a simple yet effective blocking scheme that is supported by mathematical intuition, yet remains easy to implement and efficient in practice.

\begin{algorithm}[h]
\caption{RAS-Type Temporal Blocking}
\label{alg:ras_temporal}
\SetKwInOut{Input}{Input}
\SetKwInOut{Output}{Output}

\Input{Field buffers $v_{\text{read}}, v_{\text{write}}$; tile size $(T_I,T_J)$; maximum iterations $N_{\max}$;\\inner iterations $T_{\text{inner}};$}
\Output{Updated field $v$}

$N_{\text{outer}} := \left\lceil N_{\max}/T_{\text{inner}} \right\rceil$\;

\For{$t := 1$ \KwTo $N_{\text{outer}}$}{
  Draw random shifts $s_i \in [0,T_I)$, $s_j \in [0,T_J)$ with periodic wrap-around\;
  Partition the interior into overlapping tiles of size $(T_I,T_J)$ using $(s_i,s_j)$\;
  \ForEach{tile $\mathcal{T}$ \textnormal{(in parallel)}}{
    \For{$\tau := 1$ \KwTo $T_{\text{inner}}$}{
      \For{$(i,j) \in \mathcal{T}$}{
        $v_{\text{write}}(i,j) := \text{JacobiStencil}\big(v_{\text{read}}, i, j\big)$\;
      }
      \tcp{local buffer ping--pong}
      Swap $v_{\text{read}} \leftrightarrow v_{\text{write}}$ 
    }
    \tcp{Each cell has a single writer but may be read by multiple tiles (benign races)}
  }
}
\end{algorithm}

\section{Implemented Algorithms}
\label{sec:algorithms}
As discussed in the previous sections, the dominant computational cost of the solver lies in the smoothing operations of the geometric multigrid method used for the velocity update of the inexact Uzawa method. Our optimisation efforts therefore focused on this kernel, where we systematically explored different implementation strategies to balance convergence properties and scalability. We ultimately implemented and tested the following five smoothers, in both Python (Numba) and C++ where applicable, using strong and weak scaling as the primary performance metrics:
\begin{enumerate}[label=(\roman*)]
    \item \textbf{RBGS Smoother} An implementation of the Red–Black Gauss–Seidel (RBGS) method. This is the only Gauss–Seidel variant tested and it is the algorithm currently used in Pyroclast. RBGS may offer better convergence properties than Jacobi-type smoothers, but at higher computational cost and complexity. Parallelism is achieved by splitting the grid points into two disjoint sets (red and black), which are updated sequentially. Fusing the $v_x$ and $v_y$ updates in this smoother is not a straight-forward process due to potential race conditions. 
    \item \textbf{Jacobi Smoother} 
    A simple implementation of a Jacobi smoother, which splits the velocity updates into separate $v_x$ and $v_y$ passes. This approach simplifies the loop structure but may incur higher memory traffic.
    \item \textbf{Fused Jacobi Smoother} A fused implementation of a Jacobi smoother. The updates for both $v_x$ and $v_y$ velocity components are fused within a single loop over the grid to improve data locality. This however requires extra logic in order to account for the staggered grid layout.
    \item \textbf{Blocked Fused Jacobi Smoother} 
    A fused Jacobi smoother that implements spatial blocking.
    \item \textbf{RAS-Jacobi Smoother} A Jacobi smoother that applies both spatial and RAS type temporal blocking.
\end{enumerate}

\newpage
\section{Correctness}
\label{sec:correctness}
To assess the robustness of our proposed RAS-type smoother, we performed a classic benchmark simulation and compared the convergence of our method with RBGS and Jacobi references. 
\subsection{Benchmark Problem}
\begin{figure}[ht!]
  \centering
  \includegraphics[width=0.95\linewidth]{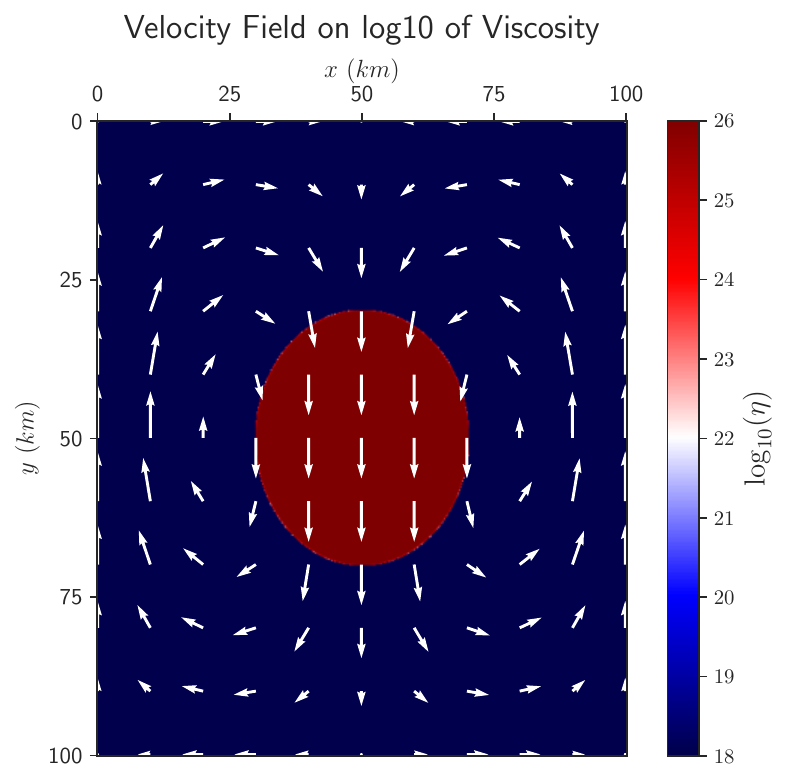}
  \caption{Velocity field (white arrows) superimposed on the logarithm of the viscosity distribution for the sinker benchmark. The spherical inclusion of viscosity $10^{26}\,\mathrm{Pa\,s}$ and radius $20\,\mathrm{km}$ sinks in a surrounding medium of viscosity $10^{18}\,\mathrm{Pa\,s}$. The inclusion descends vertically, generating a symmetric flow pattern characterized by two counter-rotating vortices in the host material.}
    
  \label{fig:material_setup}
\end{figure}

This problem consists of a highly viscous spherical inclusion sinking under gravity in a surrounding medium of lower viscosity. The difficulty of the test is controlled by the viscosity contrast between inclusion and medium: higher contrasts lead to stronger stiffness and thus increased challenges for the solver. We tested the highest viscosity contrast that is still realistic for geophysical applications. Specifically, the inclusion was assigned a viscosity of $10^{26} \ Pa \cdot s$ and a density of $3300 \ kg / m^3$, while the medium was given a viscosity of $10^{18} \ Pa \cdot s$ and a density of $3200 \ kg / m^3$. This results in a relative viscosity contrast of $10^8$. Uniform gravity was set to $g_y=10 \ m/s^2$ only in the $y$ direction. The computational domain was a square box of size $100 \times 100 \ km$ discretized at a resolution of $n_x=501$ and $n_y=601$, with the spherical inclusion of radius $20 \ km$ placed at the center. Boundary conditions were imposed to enforce physical symmetry and constrain rigid-body motion: along the left and right boundaries, $v_x$ was set to no-slip and $v_y$ to free-slip, while along the top and bottom boundaries, $v_y$ was set to no-slip and $v_x$ to free-slip. Figure~\ref{fig:material_setup} illustrates the expected velocity field: the inclusion sinks vertically with a symmetric velocity distribution that induces two characteristic vortices in the surrounding medium.
\clearpage
\subsection{Solver Configuration}
Pyroclast supports many options to tailor the behaviour of the stokes solver. For this test, we ran $1000$ Uzawa cycles with Anderson acceleration disabled. Each Uzawa cycle consisted of a single multigrid iteration, and the multigrid solver was configured with four levels and a scaling factor of $2.5$. On the finest grid, each V-cycle applied five pre-smoothing and five post-smoothing iterations. When descending to coarser levels, the number of smoothing iterations was increased by a factor of $2.5$ at each refinement step. To ensure convergence under the extreme viscosity contrast, we employed a viscosity rescaling strategy. Let $\eta_{\min} = \min \eta$ denote the minimum viscosity in the domain. At each stage, we defined a rescaled computational viscosity field
\begin{equation}
\eta_{\text{comp}} = (1 - \theta)\,\eta_{\min} + \theta\,\eta,    
\end{equation}
where the parameter $\theta$ was gradually increased from $0$ to $1$ in increments of $0.25$ at intervals of $25$ Uzawa cycles \mbox{(i.e., $\theta \in \{0,0.25,0.5,0.75,1.0\}$)}. This allows the solver to obtain increasingly better initial guesses for the solution process. The initial guess for the pressure field was set to the lithostatic distribution,
\begin{equation}
p_{\text {litho }}(x, y)=\int_0^y \rho\left(x, y^{\prime}\right) g_y d y^{\prime}
\end{equation}
which provides a physically meaningful starting point for the iteration. Since only the pressure gradient appears in the Stokes equations, the discretized system admits a null space and the pressure is determined only up to an additive constant. To remove this indeterminacy, we enforce a zero-average constraint by subtracting the domain mean from the pressure field after each update, ensuring a unique and well-defined solution.

\subsection{Smoother Setup}
\label{sec:smoother_setup}

To evaluate the performance and robustness of our smoother, we tested and compared the following configurations:

\begin{itemize}
    \item \textbf{RBGS smoother:} the classical red–black Gauss–Seidel method currently used in the codebase. It provides good convergence properties, but has higher computational cost compared to Jacobi.
    \item \textbf{Jacobi smoother:} a standard Jacobi method implemented as a simple reference. It is easier to parallelize but may be less robust compared to RBGS.
    \item \textbf{RAS smoother:} a blocked smoother based on our RAS-type temporal blocking strategy with random periodic block shifting. We use a block size of $32 \times 32$ and $T_{\text{inner}}=4$ inner iterations.
    \item \textbf{Mixed smoother:} a hybrid configuration where the finest grid level uses the Jacobi smoother, while all coarser levels use the RAS smoother.
\end{itemize}
All smoothers were tested using a velocity under-relaxation factor of $\omega_u = 0.3$ and a pressure relaxation factor of $\omega_p = 0.6$.

\subsection{Measuring Residuals}
\label{sec:residuals}

Measuring residuals in the incompressible Stokes problem requires special care, as the magnitudes of pressure and velocity differ by several orders. Since we solve the fully dimensional equations, a simple $\ell^2$ norm of the residual relative to the right-hand side is not meaningful. Instead, we compute surrogate residuals in terms of relative energy norms consistent with the physics of the problem.  

\noindent\textbf{Pressure residual.}  
For the pressure field, the natural energy norm is induced by the Schur complement $S = D M^{-1} G$ defined in Equation~\eqref{eq:schur_system}. Direct assembly of $S$ is intractable, and we already use an approximate inverse $S^{-1} \approx \widetilde{S}^{-1}$ as a positive diagonal surrogate. Using this approximation, the pressure residual can be expressed as
\begin{equation}
\begin{aligned}
\|p - p^\ast\|_{S}^{2}
&= \|S^{-1} r_p\|_{S}^{2} \\
&= (S^{-1} r_p)^{\!\top} S (S^{-1} r_p) \\
&= r_p^{\!\top} S^{-1} r_p \\
&\approx r_p^{\!\top}\,\widetilde S^{-1} r_p
= \|r_p\|_{\widetilde S^{-1}}^{2},
\end{aligned}
\end{equation}
where $p^\ast$ is the true pressure solution and $r_p$ is the continuity residual. A naive choice $\widetilde S^{-1} = \eta$ is efficient as a preconditioner, but does not account for the mesh-dependent scaling of the discrete Laplacian contained in $S$. To obtain a resolution-independent measure, we introduce the local diagonal of the discrete Laplacian into the surrogate:
\begin{equation}
\widetilde S^{-1}_{jj} = 
\frac{\eta_{jj}}{\,2/{\Delta x}^{2} + 2/{\Delta y}^{2}} \qquad \text{(in 2D)},
\end{equation}
or more generally $\widetilde S^{-1}_{jj} \approx \eta_{jj} \,(\operatorname{diag}(-\Delta_h))^{-1}_{jj}$.  
This scaling captures the $h^2$ factor in $S^{-1}$ and ensures that the resulting pressure energy residual $\|r_p\|_{\widetilde S^{-1}}$ remains consistent across different grid resolutions.

\noindent\textbf{Velocity residual.}  
The momentum operator $M$ that arises from our discretization is symmetric negative definite. To obtain a positive definite energy norm, we work with $-M$ and its diagonal surrogate \mbox{$\widetilde M^{-1} = \operatorname{diag}(-M)^{-1}$}. This is purely a matter of convention in how the PDE system is written and does not affect the residual definition. With this convention, the velocity residual becomes
\begin{equation}
\begin{aligned}
\|v - v^\ast\|_{-M}^{2}
&= \|(-M)^{-1} r_v\|_{-M}^{2} \\
&= \big((-M)^{-1} r_v\big)^{\!\top} (-M) \big((-M)^{-1} r_v\big) \\
&= r_v^{\!\top} (-M)^{-1} r_v \\
&\approx r_v^{\!\top}\,\widetilde M^{-1} r_v
= \|r_v\|_{\widetilde M^{-1}}^{2},
\end{aligned}
\end{equation}
where $v^\ast$ is the true velocity solution and $r_v$ is the velocity residual evaluated at fixed $p$ (with $Gp$ absorbed into the right-hand side).  

\noindent\textbf{Global relative energy residual.}  
Since the only non-zero right-hand side in Equation~\eqref{eq:stokes_system} is $f$, which belongs to the velocity subsystem, we normalize the global residual by the corresponding energy norm $\|f\|_{\widetilde M^{-1}}$. The surrogate relative energy residual is then given by
\begin{equation}
\frac{\|x - x^\ast\|_{-M \oplus S}}{\|x^\ast\|_{-M \oplus S}}
\;\approx\;
\sqrt{\frac{\|r_v\|_{\widetilde M^{-1}}^{2}+\|r_p\|_{\widetilde S^{-1}}^{2}}
{\|f\|_{\widetilde M^{-1}}^{2}}},
\end{equation}
where \mbox{$x=[v,p]$} denotes the approximate full state and \mbox{$x^\ast=[v^\ast,p^\ast]$} the exact solution. Here we define the block energy norm as
\begin{equation}
\|[v,p]\|_{-M \oplus S}^{2} := v^{\!\top}(-M)v \;+\; p^{\!\top} S p,
\end{equation}
with $-M$ and $S$ both symmetric positive definite under our discretisation and boundary conditions.
\clearpage
\begin{figure*}[ht!]
  \centering
  \includegraphics[width=0.89\linewidth]{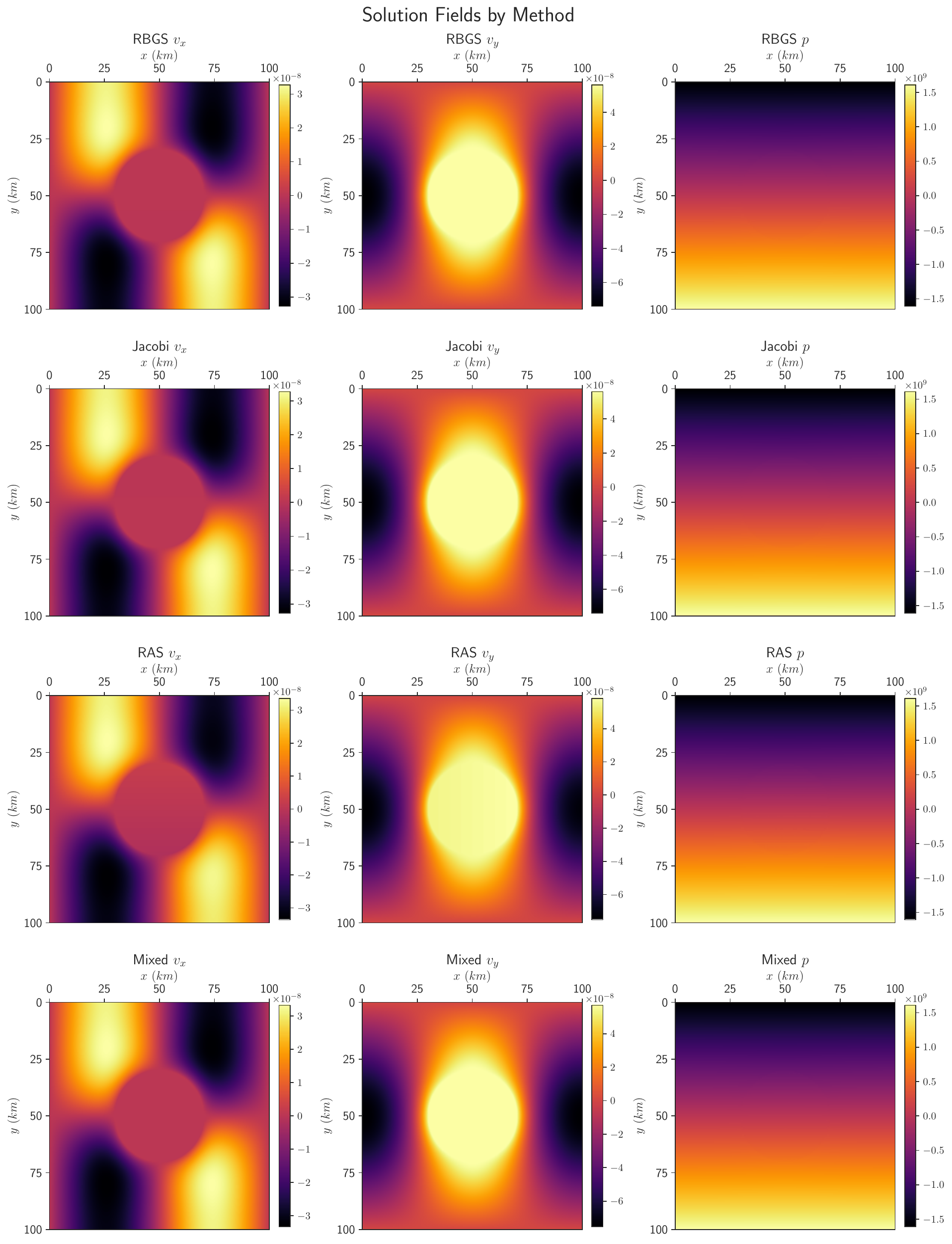}
  \caption{Velocity and pressure fields obtained with the different smoothers in the sinker benchmark. Each row corresponds to one smoother (RBGS, Jacobi, RAS, Mixed), showing the horizontal velocity $v_x$ (left), vertical velocity $v_y$ (center), and pressure $p$ (right). All methods reproduce the expected sinking behavior of the high-viscosity inclusion and the induced vortical flow in the surrounding medium. The pressure fields exhibit the correct vertical gradient associated with the density contrast. No spurious artefacts are visible, and the absolute magnitudes of $v_x$, $v_y$, and $p$ are consistent across methods, confirming that all smoothers converge to the same physical solution.}
    
  \label{fig:method_fields}
\end{figure*}
\clearpage
\noindent Hence, this formulation yields a relative energy residual that measures the fraction of the mechanical energy input by the body forces that remains unbalanced in the current iterate. The velocity contribution corresponds to unresolved viscous dissipation in the momentum equations, while the pressure contribution reflects residual divergence errors weighted by viscosity. A value of $10^{-4}$ thus indicates that only $0.01\%$ of the driving work is not captured by the numerical solution, providing a resolution-independent and physically meaningful measure of solver accuracy.
\begin{figure}[!b]
  \centering
  \includegraphics[width=\linewidth]{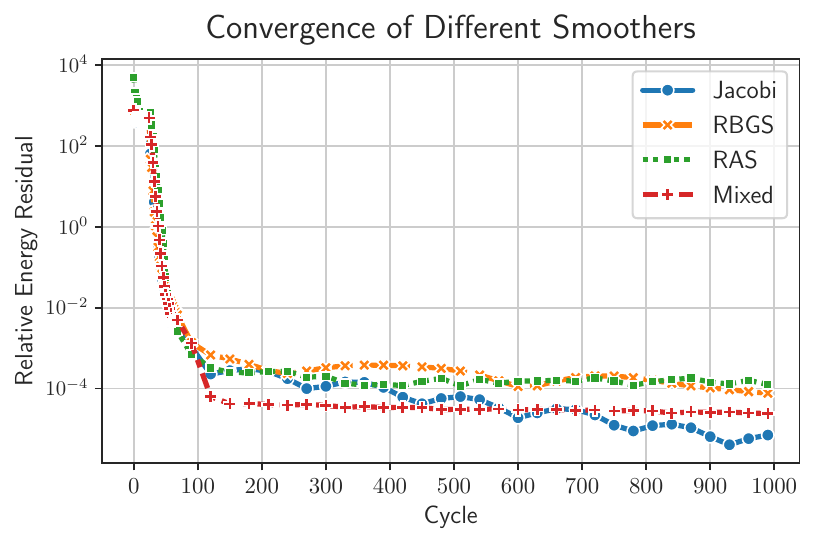}
  \caption{Convergence history of the different smoothers in the sinker benchmark. The plot shows the relative energy residual as a function of Uzawa cycles. All smoothers converge by several orders of magnitude. The Jacobi smoother achieved the lowest residual values, while the mixed smoother exhibited the most stable convergence behavior. Each curve corresponds to the worst out of ten independent runs.}
  \label{fig:residuals}
\end{figure}
\subsection{Correctness Results}
\label{sec:correctness_results}
Figure \ref{fig:residuals} presents the convergence history of the different smoothers tested in the sinker benchmark. The plot reports the relative energy residual as a function of Uzawa cycles. All smoothers demonstrate robust convergence, with residuals decreasing by several orders of magnitude. The overall performance of the methods is comparable. Among the tested smoothers, the Jacobi variant reached the lowest final residual values for this benchmark, whereas the mixed smoother exhibited the most stable and monotone convergence trajectory across the full range of cycles. Despite the application of velocity under-relaxation, oscillatory behavior can still be observed in the residual trajectories of all smoothers except the mixed configuration, which consistently suppresses such fluctuations.  It is important to emphasize that the sinker benchmark is an intrinsically unstable and chaotic test case. Small perturbations, such as variations in thread scheduling during smoothing, restriction, or prolongation, are sufficient to alter the residual trajectories. To reflect this variability, we executed ten independent runs for each smoother and report here the worst observed run. This conservative choice highlights the robustness of the algorithms even under less favorable execution conditions.  
We also note that convergence tends to slow significantly after the initial rapid drop in residuals, consistent with the well-known difficulty of solving high-contrast Stokes systems (cfr. \textit{``multi-multigrid"} in~\cite{Gerya_multigrid}). Despite these challenges, all smoothers ultimately reach residuals on the order of $10^{-4}$ or lower, confirming that the implementations are correct and capable of handling very strong viscosity contrasts. Figure~\ref{fig:method_fields} compares the velocity and pressure fields obtained with the different smoother variants. Visually, all solutions appear correct: the horizontal and vertical velocity components are smooth and consistent across methods, with no discernible artefacts or spurious oscillations. The sinking motion of the high-viscosity inclusion is captured symmetrically, and the induced vortical flow in the surrounding medium is reproduced in each case. Likewise, the pressure fields develop the expected vertical gradient, reflecting the imposed density contrast, and no anomalous structures are present. In addition to qualitative agreement, the absolute magnitudes of $v_x$, $v_y$, and $p$ are consistent across all methods, confirming that the different smoothers converge to the same physical solution. These results demonstrate that all tested smoothers yield physically reliable velocity and pressure fields despite differences in convergence behavior.

\section{Synthetic Benchmarks}
\label{sec:benchmarks}
We assess the performance of the smoother implementations introduced in Section~\ref{sec:algorithms} using synthetic benchmarks. These controlled tests isolate the stencil kernels from application-level effects, allowing us to quantify scalability, efficiency, and the impact of different optimisation strategies on modern hardware. 

For the benchmark study, we evaluated the five smoothers introduced in Section~\ref{sec:algorithms}, with implementations in C++ for all cases and in Python (Numba) for the RBGS and Jacobi variants. This allows a direct comparison between JIT-compiled Python and optimized C++ code. The RAS-Jacobi smoother was configured with a block size of $32 \times 32$ and $T_{\text{inner}} = 4$ local iterations. Unless otherwise stated, performance results are reported relative to the C++ Jacobi smoother, chosen as a simple, well-understood, and scalable baseline. All implementations were validated against the Pyroclast framework, with further verification provided by the benchmark problem described in Section~\ref{sec:correctness}.

\subsection{Benchmark Setup}
\label{sec:benchmark_setup}
All synthetic benchmark experiments are conducted on a two-dimensional problem discretised on a uniform rectangular grid. For strong scaling, we performed three distinct experiments with grid sizes of $2000 \times 2000$, $8000 \times 8000$, and $15000 \times 15000$ points. For weak scaling, we began with a $2000 \times 2000$ problem on four cores and increased the grid dimensions by a factor of $\sqrt{2}$ in both directions for each doubling of the available computational resources, ensuring constant workload per core. Starting from four cores avoids unrealistically optimistic cache performance that would otherwise skew results on smaller problem sizes. Each run consisted of one warm-up iteration followed by twenty smoothing iterations. This procedure was repeated ten times, and we report the median runtime together with the $95\%$ confidence interval of the median, following the recommendations of~\cite{scientific_benchmarking}.

\subsection{System Configuration}
\label{sec:system_configuration}
All experiments were performed on the Santis cluster, part of the ALPS supercomputer at the Swiss National Supercomputing Centre (CSCS). Each Santis node integrates four NVIDIA GH200 Grace Hopper superchips interconnected by cache-coherent NVLink. Every Grace CPU provides 72 ARM cores that share a single NUMA domain, while scaling beyond 72 threads requires communication across NVLink between sockets. Each superchip is equipped with 120~GB of LPDDR5X CPU memory and 94~GB of HBM3 GPU memory, and the four-way NVLink configuration ensures high-bandwidth data movement between CPUs and GPUs. A detailed discussion on the architecture of the ALPS GH200 nodes is given in~\cite{fusco2024understandingdatamovementtightly}. This topology is important to consider when interpreting performance at higher core counts. 

For the synthetic benchmarks, we restricted execution to a single node and a single MPI rank, making use of up to 256 CPU cores across the four Grace CPUs. The software environment was based on SUSE Linux Enterprise Server 15 SP5 with GCC 13.3.0 as the C++ compiler. Python experiments used Python 3.12.5 with Numba 0.61.2 for JIT compilation. A summary of the hardware and software configuration is provided in Table~\ref{tab:system_specs}.
\begin{table}
\centering
\begin{tabular}{ll}
\hline
\multicolumn{2}{l}{\textit{\textbf{Hardware per Socket (4x per node)}}} \\
Superchip      & NVIDIA GH200 Grace Hopper \\
Memory  & 120~GB LPDDR5X (CPU) + 94~GB HBM3 (GPU) \\
CPU & Grace CPU (72-core ARM) \\
GPU & Hopper GPU \\
Cache (L1 to L3)  & 4.5 MiB (I) + 4.5 MiB (D), 72 MiB, 114 MiB \\
\hline
\multicolumn{2}{l}{\textit{\textbf{Software}}} \\
Operating System & SUSE Linux Enterprise Server 15 SP5 \\
C++ Compiler     & GCC 13.3.0 \\
Python         & Python 3.12.5 \\
Numba          & Numba 0.61.2 \\
\hline
\end{tabular}
\caption{Hardware and software configuration of a GH200 node of the ALPS supercomputer.}
\label{tab:system_specs}
\end{table}
\subsection{Strong Scaling Results}
\label{sec:strong_scaling}
\Cref{fig:strong_scaling_ALPS} reports the strong scaling results achieved by the different smoothers. Speedups are reported relative to the C++ implementation of the classic Jacobi smoother, which we use as a baseline since it is a simple, well-understood, and scalable standard.

A first observation is that the Numba JIT-compiled smoothers consistently underperform compared to their C++ counterparts. While Numba provides a significant improvement over pure Python, it cannot match the efficiency of carefully optimized C++ implementations on this architecture. 
Across all problem sizes, the lowest runtimes were achieved by the RAS-Jacobi smoother at 64 cores. Beyond this point, performance deteriorates for all algorithms due to the need for data movement across sockets via NVLink. Nevertheless, the RAS-Jacobi smoother remains markedly faster than the other approaches at both 128 and 256 cores. This suggests that extending the method to 4D temporal blocking for a 3D Stokes problem could enable effective scaling beyond a single NUMA domain.  
In general, the advantage of the RAS-Jacobi smoother becomes more pronounced at higher core counts. Empirically, this can be attributed to cache behaviour: when scaling to more cores, all smoothers suffer from increased cache misses, but the RAS-Jacobi smoother achieves consistently better cache hit rates. On the smallest grid ($2000 \times 2000$), RAS-type temporal blocking only yields benefits once at least 32 cores are employed, reflecting the overhead of blocking for very small problems. If we examine the speedup curves of the RAS-Jacobi smoother, we find that its peak relative speedup occurs at 128 cores. This is not due to a further improvement of the RAS-Jacobi implementation itself, but rather because the baseline Jacobi smoother experiences a sharp drop in performance once computation spans two NUMA domains. At 256 cores, performance improves slightly again despite involving four NUMA domains, which we attribute to more balanced scheduling: in effect, one “fast” and one “slow” NUMA domain is less balanced than one fast domain combined with three slower ones.
An interesting effect is observed when comparing problem sizes. Relative to the Jacobi baseline, RAS-Jacobi achieves around a $3\times$ speedup for the $8000 \times 8000$ case but only about a $2\times$ speedup for the largest $15000 \times 15000$ case. This reduction is likely linked to the higher rate of TLB misses arising from the memory access patterns in very large two-dimensional arrays. In such cases, the cost of address translation becomes significant. One potential remedy would be to copy tile-local data into small ping–pong buffers and carry out the inner iterations there, thereby also eliminating the benign race condition discussed in Algorithm~\ref{alg:ras_temporal}. However, this would also introduce redundant halo loads, increasing memory traffic. 
Finally, while the performance of the RAS-Jacobi smoother can be trivially improved by increasing the number of inner iterations, we find that going beyond four inner steps actually slows down convergence of the overall solver. This highlights the trade-off between per-iteration performance and solver-level convergence efficiency.

Overall, the results confirm that RAS-type temporal blocking offers the best balance of performance and scalability among the tested smoothers.
\subsection{Weak Scaling Results}
\label{sec:weak_scaling}
\begin{figure*}[ht!]
  \centering
  \begin{subfigure}{\linewidth}

    \centering
    \includegraphics[width=0.85\linewidth]{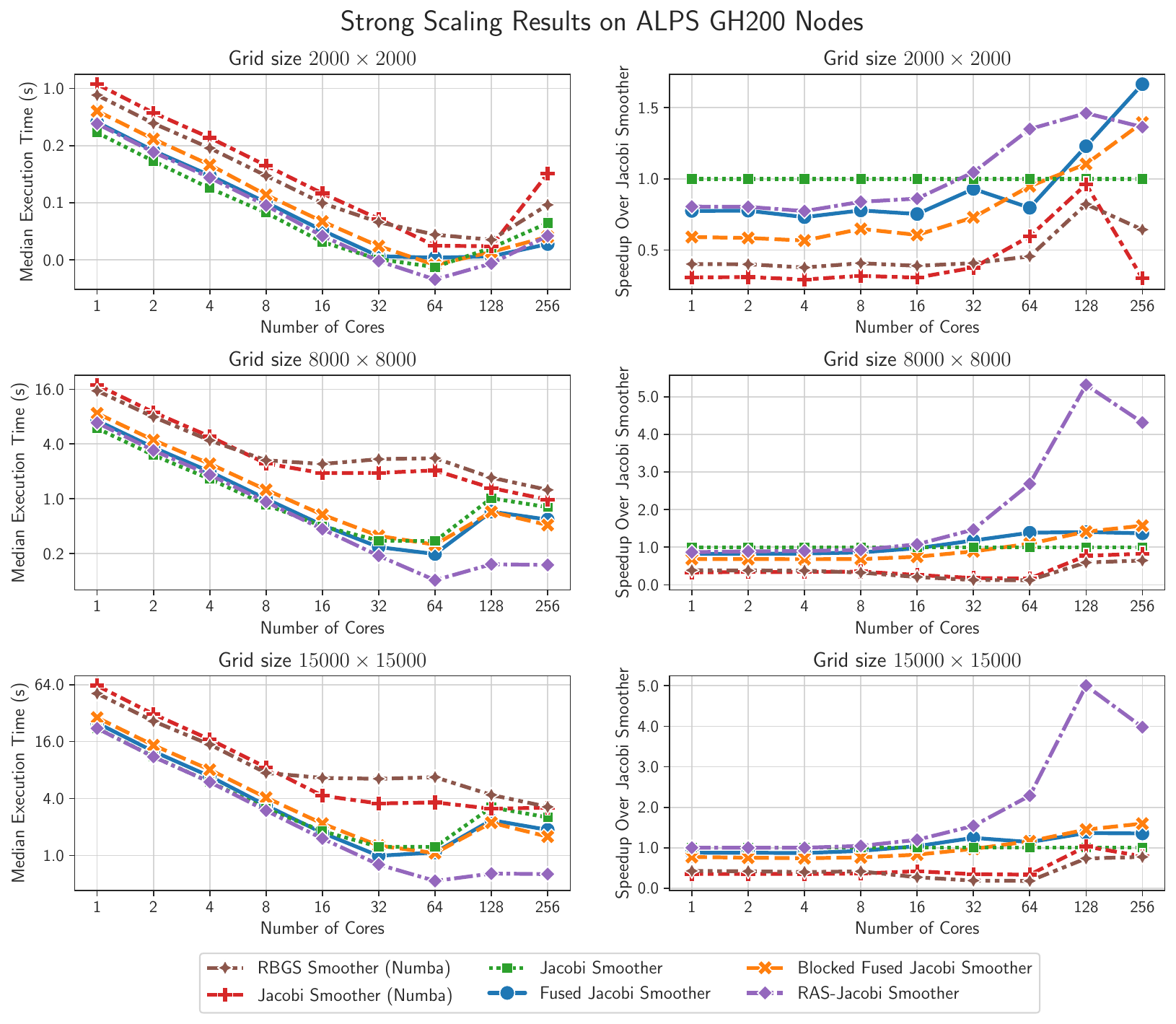}
    \caption{Strong scaling of the smoothers evaluated on Alps GH200 nodes. The left column shows the median execution time for three distinct grid sizes: $2000 \times 2000$, $8000 \times 8000$, and $15000 \times 15000$. The right column illustrates the corresponding speedup, calculated relative to the median execution of the C++ Jacobi smoother.}
    \label{fig:strong_scaling_ALPS}
  \end{subfigure}

  \vspace{1em} 

  \begin{subfigure}{\linewidth}
    \centering
    \includegraphics[width=0.85\linewidth]{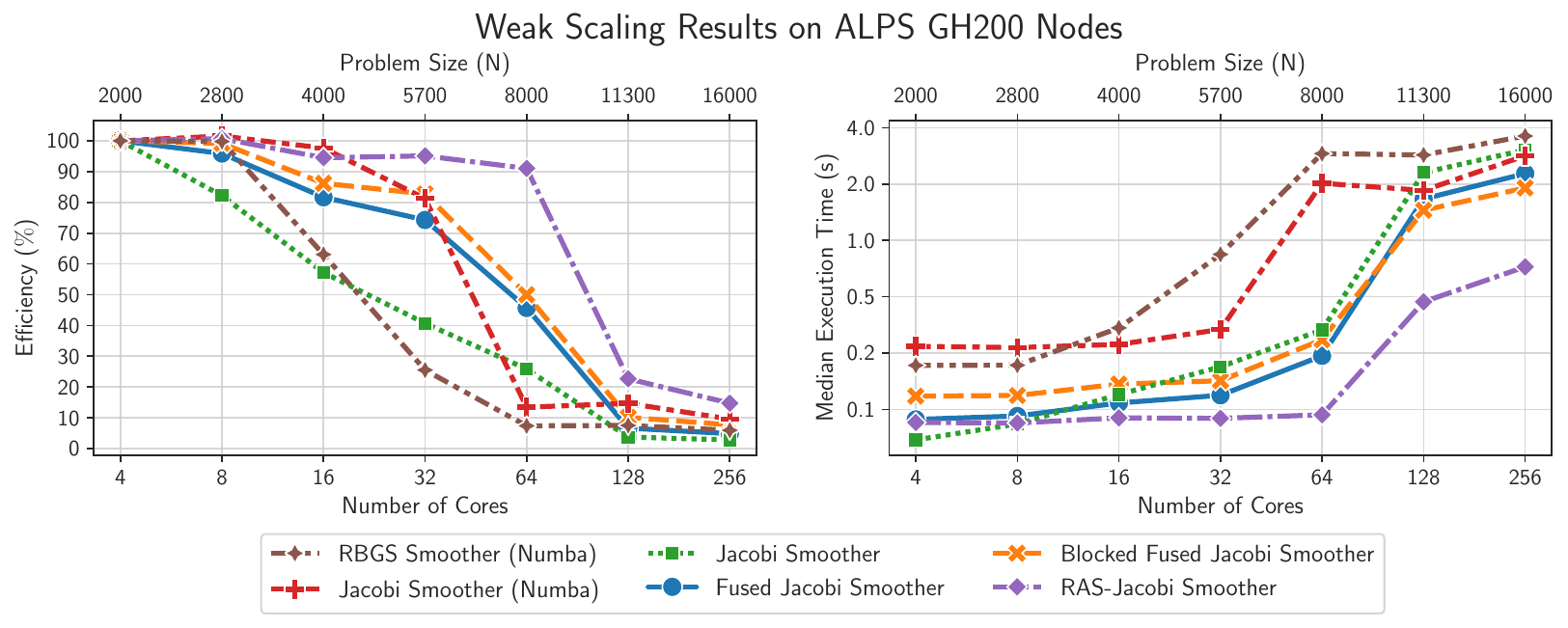}
    \caption{Weak scaling of the smoothers evaluated on Alps GH200 nodes. As the number of threads increases to 256, the problem size is scaled proportionally to maintain a constant workload per thread. The left plot shows the scaling efficiency, calculated relative to a base grid size of $2000 \times 2000$ points. The right plot shows the median execution time over 10 repetitions.}
    \label{fig:weak_scaling_ALPS}
  \end{subfigure}

  \caption{Strong and weak scaling of the smoothers evaluated on Alps GH200 nodes. Subfigure~\subref{fig:strong_scaling_ALPS} shows strong scaling results, while Subfigure~\subref{fig:weak_scaling_ALPS} shows weak scaling results.}
  \label{fig:scaling_ALPS}
\end{figure*}

\Cref{fig:weak_scaling_ALPS} presents the weak scaling behaviour of the six smoother implementations. We report both efficiency, defined as $E = T_{4\ \text{cores}} / T_{n\ \text{cores}}$, and the corresponding median execution times. The results indicate that, except for the smallest case, the RAS-Jacobi smoother consistently delivered the best performance across all configurations. For the $2000 \times 2000$ grid on four cores, the classic Jacobi smoother achieved the lowest runtime. The efficiency plots show that although all smoothers experience a steady decline in scaling efficiency, the RAS-Jacobi smoother sustains over 90\% efficiency up to 64 cores. Beyond this point, efficiency drops sharply as computation crosses NUMA domains, reflecting the additional overhead of data transfers across NVLink interconnects. Comparing the C++ Jacobi variants, we find that the plain, fused, and blocked fused implementations perform almost identically, both in terms of efficiency and runtime. This indicates that simple loop fusion and blocking does not improve the weak scaling behaviour for a 2D problem. By contrast, the RAS-Jacobi smoother retains a significant performance edge, particularly at higher thread counts, demonstrating the benefits of temporal blocking under high memory bandwidth and cache pressure. Finally, the Numba-based smoothers (RBGS and Jacobi) show very poor weak scaling: their efficiency collapses rapidly with increasing core count, and their runtimes diverge significantly from the C++ implementations at scale. As previously mentiond, RAS type domain decomposition has long been used in distributed-memory settings to improve weak scaling and reduce communication overhead. It is interesting to observe its effectiveness also at the socket and NUMA level, demonstrating that the same principle can yield tangible benefits at a much smaller scale, where the bottleneck shifts from inter-node communication to cache and memory hierarchy effects.
\clearpage
\clearpage
\section{Conclusions}
\label{sec:conclusions}
We presented the design, implementation, and evaluation of optimized matrix-free stencil kernels for multigrid smoothing in the variable-viscosity Stokes equations. We investigated five smoother variants derived from different optimisation strategies, ranging from classical Red–Black Gauss–Seidel and Jacobi methods to fused, blocked, and RAS-type temporally blocked implementations. To ensure correctness, we validated our approach using the high-contrast sinker benchmark and introduced an energy-based residual norm that balances velocity and pressure contributions, enabling reliable convergence assessment. Our performance study on NVIDIA GH200 Grace Hopper nodes of the ALPS supercomputer demonstrates that while all smoothers scale well within a single NUMA domain, the RAS-Jacobi smoother consistently delivers the best performance at higher core counts. It sustains over $90\%$ weak-scaling efficiency up to 64 cores and achieves up to a threefold speedup compared to the classic Jacobi baseline, owing to its optimized cache usage, especially when scaling to multiple parallel cores. These results show that RAS-type blocking, already used in distributed-memory solvers to reduce communication, can also yield substantial benefits at the socket and NUMA level. Taken together, our results highlight that carefully designed, cache-aware stencil kernels are essential for extracting performance from modern heterogeneous architectures.
\section{Future Work}
\label{sec:future_work}
The results of this study suggest several promising directions for future research. First, given the comparatively strong scaling behaviour of the RAS-type smoother in 2D across NUMA domains, it would be natural to extend this strategy to 3D problems. This would effectively enable a 4D blocking scheme and could provide further improvements in cache reuse and parallel efficiency. Moreover, 3D geodynamic problems are typically better conditioned than their two-dimensional counterparts, making it important to assess how convergence is affected in practice. Scaling to multiple NUMA domains also becomes more relevant in three dimensions, where larger problem sizes may help to amortize the overhead of inter-socket communication. We therefore expect that the RAS-type smoother may unlock more efficient scaling across all cores of a multi-socket node in this setting. Finally, a natural extension of this work is to GPUs. A similar temporal blocking strategy could be applied at the GPU level by operating on larger patches and preloading the required stencil data for example into shared memory close to the streaming multiprocessors. Exploring such techniques may provide a path toward cache-efficient stencil kernels that fully exploit the memory hierarchy of modern heterogeneous architectures.

\section*{Acknowledgements}
We would like to thank Prof.~Dr.~Oliver Fuhrer (MeteoSwiss, ETH Zurich) for the insightful discussions and exchange of ideas that helped shape this work.

\noindent We also gratefully acknowledge the Swiss National Supercomputing Centre (CSCS) for providing the computing resources that made the experiments in this study possible.

\section*{Code Availability}

The Pyroclast framework, which forms the basis of this work, is openly developed at  
\url{https://github.com/MarcelFerrari/Pyroclast}.  
The project is still in an early stage, but it is actively maintained and regularly updated.

\FloatBarrier

\bibliographystyle{acm}
\bibliography{Bibliography/Bibliography}
\FloatBarrier

\onecolumn
\section*{Appendix}
\begin{table}[h]
\centering
\caption{Weak scaling performance of the smoothers on a single, shared-memory Alps GH200 Node. The simulation maintains a constant problem size of 1000x1000 per thread, meaning the total grid size increases as more threads are used. The simulation was run with a maximum of 20 iterations and a memory alignment of 64 bytes. Inner temporal blocking and tiling were both set to 4. The results are based on 10 repetitions, and the table reports the median, 95\% confidence interval (CI) lower and upper bounds, and the speedup relative to the single-thread median. The scaling was tested for up to 256 threads.}\label{tab:weak_scaling_10}

\begin{tabular}{l l r r r r r r r r r}
\toprule
\multirow{2}{*}{Algorithm} & \multirow{2}{*}{} & \multicolumn{9}{c}{Threads} \\
\cmidrule{3-11}
 & & 1 & 2 & 4 & 8 & 16 & 32 & 64 & 128 & 256 \\
\midrule
\multirow{4}{*}{Jacobi Smoother (Numba)} & Median (s) & 0.26 & 0.26 & 0.27 & 0.26 & 0.28 & 0.29 & 1.24 & 1.29 & 2.33 \\
 & 95\% CI Lower (s) & 0.26 & 0.26 & 0.27 & 0.26 & 0.28 & 0.29 & 0.94 & 1.17 & 2.04 \\
 & 95\% CI Upper (s) & 0.26 & 0.26 & 0.27 & 0.27 & 0.28 & 0.29 & 1.29 & 1.33 & 2.37 \\
 & Speedup & 1.00 & 1.02 & 0.99 & 1.00 & 0.96 & 0.91 & 0.21 & 0.20 & 0.11 \\
\midrule
\multirow{4}{*}{RBGS Smoother (Numba)} & Median (s) & 0.25 & 0.21 & 0.22 & 0.22 & 0.34 & 0.84 & 2.91 & 2.86 & 3.61 \\
 & 95\% CI Lower (s) & 0.25 & 0.21 & 0.21 & 0.22 & 0.34 & 0.76 & 2.48 & 2.63 & 3.35 \\
 & 95\% CI Upper (s) & 0.25 & 0.21 & 0.22 & 0.22 & 0.35 & 1.45 & 3.11 & 3.15 & 3.78 \\
 & Speedup & 1.00 & 1.22 & 1.17 & 1.17 & 0.74 & 0.30 & 0.09 & 0.09 & 0.07 \\
 \midrule

\multirow{4}{*}{Jacobi Smoother} & Median (s) & 0.08 & 0.08 & 0.09 & 0.10 & 0.15 & 0.21 & 0.33 & 2.30 & 3.05 \\
 & 95\% CI Lower (s) & 0.08 & 0.08 & 0.09 & 0.10 & 0.15 & 0.21 & 0.33 & 2.15 & 2.78 \\
 & 95\% CI Upper (s) & 0.08 & 0.08 & 0.09 & 0.11 & 0.15 & 0.21 & 0.34 & 2.45 & 3.22 \\
 & Speedup & 1.00 & 1.01 & 0.92 & 0.76 & 0.53 & 0.38 & 0.24 & 0.03 & 0.03 \\
\midrule

\multirow{4}{*}{Fused Jacobi Smoother} & Median (s) & 0.10 & 0.10 & 0.11 & 0.12 & 0.14 & 0.15 & 0.24 & 1.67 & 2.28 \\
 & 95\% CI Lower (s) & 0.10 & 0.10 & 0.11 & 0.11 & 0.13 & 0.15 & 0.24 & 1.53 & 2.13 \\
 & 95\% CI Upper (s) & 0.10 & 0.10 & 0.11 & 0.12 & 0.14 & 0.15 & 0.24 & 1.78 & 2.33 \\
 & Speedup & 1.00 & 1.00 & 0.91 & 0.87 & 0.74 & 0.67 & 0.41 & 0.06 & 0.04 \\
\midrule

\multirow{4}{*}{Blocked Fused Jacobi Smoother} & Median (s) & 0.11 & 0.11 & 0.15 & 0.15 & 0.17 & 0.18 & 0.29 & 1.45 & 1.91 \\
 & 95\% CI Lower (s) & 0.11 & 0.11 & 0.15 & 0.15 & 0.17 & 0.18 & 0.29 & 1.39 & 1.85 \\
 & 95\% CI Upper (s) & 0.11 & 0.11 & 0.15 & 0.15 & 0.18 & 0.18 & 0.30 & 1.59 & 1.98 \\
 & Speedup & 1.00 & 0.97 & 0.74 & 0.74 & 0.64 & 0.62 & 0.37 & 0.08 & 0.06 \\
\midrule
\multirow{4}{*}{RAS Jacobi Smoother} & Median (s) & 0.09 & 0.09 & 0.11 & 0.11 & 0.11 & 0.11 & 0.12 & 0.47 & 0.72 \\
 & 95\% CI Lower (s) & 0.09 & 0.09 & 0.11 & 0.11 & 0.11 & 0.11 & 0.12 & 0.41 & 0.69 \\
 & 95\% CI Upper (s) & 0.09 & 0.09 & 0.11 & 0.11 & 0.11 & 0.11 & 0.12 & 0.51 & 0.73 \\
 & Speedup & 1.00 & 0.99 & 0.86 & 0.86 & 0.81 & 0.81 & 0.78 & 0.19 & 0.13 \\

\bottomrule
\end{tabular}
\end{table}

\begin{table}[tp]
\centering

\caption{Strong scaling performance of the smoothers on a single, shared-memory Alps GH200 Node with a 15000x15000 grid size. The simulation was run with a maximum of 20 iterations and a memory alignment of 64 bytes. Inner temporal blocking and tiling were set to 4. The results are based on 10 repetitions, and the table reports the median, 95\% confidence interval (CI) lower and upper bounds, and the speedup relative to the single-thread median. The scaling was tested for up to 256 threads.}
\label{tab:strong_scaling_run_8}
\begin{tabular}{l l r r r r r r r r r}
\toprule
\multirow{2}{*}{Algorithm} & \multirow{2}{*}{} & \multicolumn{9}{c}{Threads} \\
\cmidrule{3-11}
 & & 1 & 2 & 4 & 8 & 16 & 32 & 64 & 128 & 256 \\
\midrule

\multirow{4}{*}{Jacobi Smoother (Numba)} & Median (s) & 61.03 & 30.57 & 16.64 & 8.39 & 4.17 & 2.13 & 2.17 & 2.10 & 2.57 \\
 & 95\% CI Lower (s) & 60.99 & 30.53 & 16.60 & 8.34 & 4.16 & 2.12 & 2.09 & 1.95 & 2.38 \\
 & 95\% CI Upper (s) & 61.10 & 30.67 & 16.67 & 8.43 & 4.29 & 2.14 & 2.30 & 2.21 & 3.03 \\
 & Speedup & 1.00 & 2.00 & 3.67 & 7.27 & 14.64 & 28.66 & 28.11 & 29.06 & 23.71 \\
\midrule
\multirow{4}{*}{RBGS Smoother (Numba)} & Median (s) & 51.55 & 26.29 & 14.88 & 7.46 & 6.61 & 6.47 & 6.73 & 4.37 & 3.27 \\
 & 95\% CI Lower (s) & 51.14 & 25.97 & 14.27 & 7.32 & 6.38 & 6.27 & 6.52 & 4.26 & 2.86 \\
 & 95\% CI Upper (s) & 51.74 & 26.50 & 15.03 & 7.62 & 6.77 & 6.57 & 6.82 & 4.65 & 3.33 \\
 & Speedup & 1.00 & 1.96 & 3.46 & 6.91 & 7.80 & 7.97 & 7.66 & 11.79 & 15.77 \\
 \midrule
 \multirow{4}{*}{Jacobi Smoother} & Median (s) & 22.13 & 11.07 & 6.00 & 3.13 & 1.81 & 1.24 & 1.23 & 3.23 & 2.53 \\
 & 95\% CI Lower (s) & 21.92 & 11.04 & 5.97 & 3.11 & 1.79 & 1.22 & 1.22 & 3.18 & 2.46 \\
 & 95\% CI Upper (s) & 22.34 & 11.17 & 6.02 & 3.15 & 1.82 & 1.25 & 1.24 & 3.29 & 2.56 \\
 & Speedup & 1.00 & 2.00 & 3.69 & 7.07 & 12.23 & 17.87 & 17.95 & 6.86 & 8.74 \\
\midrule
 \multirow{4}{*}{Fused Jacobi Smoother} & Median (s) & 25.16 & 12.61 & 6.88 & 3.39 & 1.74 & 0.99 & 1.08 & 2.37 & 1.87 \\
 & 95\% CI Lower (s) & 25.07 & 12.55 & 6.82 & 3.37 & 1.73 & 0.99 & 1.07 & 2.30 & 1.68 \\
 & 95\% CI Upper (s) & 25.22 & 12.66 & 6.94 & 3.41 & 1.78 & 1.00 & 1.09 & 2.45 & 1.90 \\
 & Speedup & 1.00 & 2.00 & 3.65 & 7.43 & 14.47 & 25.31 & 23.31 & 10.63 & 13.45 \\
\midrule
\multirow{4}{*}{Blocked Fused Jacobi Smoother} & Median (s) & 28.65 & 14.64 & 8.07 & 4.09 & 2.18 & 1.27 & 1.06 & 2.23 & 1.59 \\
 & 95\% CI Lower (s) & 28.61 & 14.62 & 8.02 & 4.08 & 2.17 & 1.27 & 1.05 & 2.18 & 1.46 \\
 & 95\% CI Upper (s) & 28.70 & 14.67 & 8.11 & 4.12 & 2.19 & 1.28 & 1.07 & 2.24 & 1.64 \\
 & Speedup & 1.00 & 1.96 & 3.55 & 7.00 & 13.16 & 22.50 & 26.98 & 12.87 & 18.07 \\
\midrule
\multirow{4}{*}{RAS Jacobi Smoother} & Median (s) & 22.07 & 11.05 & 5.98 & 2.99 & 1.51 & 0.80 & 0.54 & 0.65 & 0.64 \\
 & 95\% CI Lower (s) & 22.02 & 11.04 & 5.94 & 2.98 & 1.51 & 0.80 & 0.53 & 0.64 & 0.63 \\
 & 95\% CI Upper (s) & 22.14 & 11.05 & 6.03 & 3.00 & 1.52 & 0.81 & 0.56 & 0.67 & 0.67 \\
 & Speedup & 1.00 & 2.00 & 3.69 & 7.39 & 14.60 & 27.49 & 40.91 & 34.20 & 34.64 \\
\bottomrule
\end{tabular}
\end{table}

\begin{table}[tp]
\centering
\caption{Strong scaling performance of the smoothers on a single, shared-memory Alps GH200 Node with a 8000x8000 grid size. The simulation was run with a maximum of 20 iterations and a memory alignment of 64 bytes. Inner temporal blocking and tiling were set to 4. The results are based on 10 repetitions, and the table reports the median, 95\% confidence interval (CI) lower and upper bounds, and the speedup relative to the single-thread median. The scaling was tested for up to 256 threads.}
\label{tab:strong_scaling_run_7}
\begin{tabular}{l l r r r r r r r r r r r r}
\toprule
\multirow{2}{*}{Algorithm} & \multirow{2}{*}{} & \multicolumn{9}{c}{Threads} \\
\cmidrule{3-11}
 & & 1 & 2 & 4 & 8 & 16 & 32 & 64 & 128 & 256 \\
 \midrule
\multirow{4}{*}{RBGS Smoother (Numba)} & Median (s) & 15.38 & 7.94 & 4.38 & 2.65 & 2.41 & 2.73 & 2.80 &  1.71  & 1.25 \\
 & 95\% CI Lower (s) & 14.74 & 7.73 & 4.23 & 2.52 & 2.32 & 2.43 & 2.46 & 1.45& 1.03 \\
 & 95\% CI Upper (s) & 15.96 & 8.19 & 4.39 & 2.93 & 2.84 & 3.05 & 3.18 & 1.90 & 1.33 \\
 & Speedup & 1.00 & 1.94 & 3.51 & 5.80 & 6.37 & 5.64 & 5.50 & 8.98 & 12.27 \\ 
\midrule
\multirow{4}{*}{Jacobi Smoother (Numba)} & Median (s) & 17.54 & 8.81 & 4.79 & 2.38 & 1.23 & 1.20 & 1.36 & 0.87 & 0.73 \\
 & 95\% CI Lower (s) & 17.40 & 8.73 & 4.72 & 2.37 & 1.22 & 0.91 & 1.17 & 0.69 & 0.73 \\
 & 95\% CI Upper (s) & 17.58 & 8.88 & 4.89 & 2.46 & 1.24 & 1.29 & 1.43 & 0.93 & 0.74 \\
 & Speedup & 1.00 & 1.99 & 3.66 & 7.36 & 14.30 & 14.64 & 12.86 & 20.21 & 24.06 \\
\midrule
\multirow{4}{*}{Jacobi Smoother } & Median (s) & 5.95 & 3.02 & 1.65 & 0.86 & 0.50 & 0.35 & 0.34 & 1.02 & 0.81\\
 & 95\% CI Lower (s) & 5.92 & 3.02 & 1.65 & 0.85 & 0.49 & 0.35 & 0.34 & 0.94 &  0.80 \\
 & 95\% CI Upper (s) & 6.01 & 3.06 & 1.68 & 0.89 & 0.51 & 0.36 & 0.35 & 1.03 &  0.81 \\
 & Speedup & 1.00 & 1.97 & 3.60 & 6.91 & 11.89 & 17.12 & 17.35 & 5.86 &  7.34 \\
\midrule
\multirow{4}{*}{Fused Jacobi Smoother } & Median (s) & 7.25 & 3.67 & 1.99 & 1.00 & 0.51 & 0.29 & 0.25 & 0.72 & 0.59  \\
 & 95\% CI Lower (s) & 7.18 & 3.65 & 1.93 & 0.99 & 0.51 & 0.29 & 0.24 & 0.72  & 0.58  \\
 & 95\% CI Upper (s) & 7.31 & 3.70 & 2.02 & 1.02 & 0.52 & 0.30 & 0.25 & 0.73  & 0.60  \\
 & Speedup & 1.00 & 1.98 & 3.64 & 7.25 & 14.09 & 24.59 & 29.28 & 10.02 & 12.24 \\
\midrule
\multirow{4}{*}{Blocked Fused Jacobi Smoother} & Median (s) & 8.69 & 4.42 & 2.43 & 1.26 & 0.67 & 0.39 & 0.31 & 0.72 &  0.521 \\
 & 95\% CI Lower (s) & 8.60 & 4.40 & 2.37 & 1.25 & 0.67 & 0.39 & 0.31 & 0.72 & 0.51  \\
 & 95\% CI Upper (s) & 8.74 & 4.44 & 2.44 & 1.27 & 0.67 & 0.40 & 0.32  & 0.72 &  0.53 \\
 & Speedup & 1.00 & 1.97 & 3.58 & 6.92 & 12.98 & 22.22 & 27.73 & 12.11 & 16.86 \\
\midrule
\multirow{4}{*}{RAS Jacobi Smoother} & Median (s) & 6.85 & 3.40 & 1.83 & 0.93 & 0.46 & 0.24 & 0.13 & 0.19 & 0.19 \\
 & 95\% CI Lower (s) & 6.81 & 3.40 & 1.83 & 0.92 & 0.46 & 0.24 & 0.13 & 0.19 &  0.19  \\
 & 95\% CI Upper (s) & 6.88 & 3.42 & 1.84 & 0.93 & 0.47 & 0.24 & 0.13 & 0.19 &  0.19  \\
 & Speedup & 1.00 & 2.01 & 3.73 & 7.39 & 14.74 & 28.95 & 53.67 & 35.85 & 36.46  \\
\bottomrule
\end{tabular}
\end{table}

\begin{table}[tp]
\centering
\caption{Strong scaling performance of the smoothers on a single, shared-memory Alps GH200 Node with a 2000x2000 grid size. The simulation was run with a maximum of 20 iterations and a memory alignment of 64 bytes. Inner temporal blocking and tiling were set to 4. The results are based on 10 repetitions, and the table reports the median, 95\% confidence interval (CI) lower and upper bounds, and the speedup relative to the single-thread median. The scaling was tested for up to 256 threads.}
\label{tab:strong_scaling_run_6}
\begin{tabular}{l l r r r r r r r r r r r r}
\toprule
\multirow{2}{*}{Algorithm} & \multirow{2}{*}{} & \multicolumn{9}{c}{Threads} \\
\cmidrule{3-11}
 & & 1 & 2 & 4 & 8 & 16 & 32 & 64 & 128 & 256 \\
\midrule
\multirow{4}{*}{RBGS Smoother (Numba)} & Median (s) & 0.85 & 0.43 & 0.24 & 0.12 & 0.06 & 0.04 & 0.03 & 0.03 & 0.06 \\
 & 95\% CI Lower (s) & 0.85 & 0.43 & 0.23 & 0.12 & 0.06 & 0.04 & 0.03 & 0.02 & 0.06 \\
 & 95\% CI Upper (s) & 0.86 & 0.43 & 0.24 & 0.12 & 0.07 & 0.04 & 0.03 & 0.03 & 0.06 \\
 & Speedup & 1.00 & 1.98 & 3.61 & 7.06 & 13.75 & 21.80 & 29.48 & 33.38 & 14.32 \\
\midrule
\multirow{4}{*}{Jacobi Smoother (Numba)} & Median (s) & 1.08 & 0.54 & 0.30 & 0.15 & 0.08 & 0.04 & 0.02 & 0.03 & 0.12 \\
 & 95\% CI Lower (s) & 1.08 & 0.54 & 0.29 & 0.15 & 0.07 & 0.04 & 0.02 & 0.03 & 0.12 \\
 & 95\% CI Upper (s) & 1.08 & 0.54 & 0.30 & 0.16 & 0.08 & 0.05 & 0.03 & 0.03 & 0.12 \\
 & Speedup & 1.00 & 1.99 & 3.64 & 7.34 & 14.38 & 28.12 & 47.41 & 41.44 & 9.07 \\
\midrule
\multirow{4}{*}{Jacobi Smoother} & Median (s) & 0.34 & 0.17 & 0.09 & 0.05 & 0.02 & 0.02 & 0.01 & 0.02 & 0.04 \\
 & 95\% CI Lower (s) & 0.34 & 0.17 & 0.09 & 0.04 & 0.02 & 0.02 & 0.01 & 0.02 & 0.04 \\
 & 95\% CI Upper (s) & 0.34 & 0.17 & 0.10 & 0.05 & 0.02 & 0.02 & 0.02 & 0.03 & 0.04 \\
 & Speedup & 1.00 & 1.99 & 3.85 & 6.96 & 14.19 & 21.45 & 26.00 & 16.30 & 8.94 \\
\midrule
\multirow{4}{*}{Fused Jacobi Smoother} & Median (s) & 0.44 & 0.22 & 0.12 & 0.06 & 0.03 & 0.02 & 0.02 & 0.02 & 0.02 \\
 & 95\% CI Lower (s) & 0.44 & 0.22 & 0.12 & 0.06 & 0.03 & 0.02 & 0.01 & 0.02 & 0.02 \\
 & 95\% CI Upper (s) & 0.44 & 0.22 & 0.12 & 0.07 & 0.04 & 0.02 & 0.02 & 0.02 & 0.03 \\
 & Speedup & 1.00 & 2.00 & 3.63 & 6.99 & 13.79 & 25.73 & 26.74 & 25.87 & 19.20 \\
\midrule
\multirow{4}{*}{Blocked Fused Jacobi Smoother} & Median (s) & 0.58 & 0.29 & 0.16 & 0.08 & 0.04 & 0.02 & 0.01 & 0.02 & 0.03 \\
 & 95\% CI Lower (s) & 0.58 & 0.29 & 0.16 & 0.08 & 0.04 & 0.02 & 0.01 & 0.02 & 0.02 \\
 & 95\% CI Upper (s) & 0.58 & 0.29 & 0.16 & 0.08 & 0.05 & 0.03 & 0.01 & 0.02 & 0.03 \\
 & Speedup & 1.00 & 1.97 & 3.69 & 7.64 & 14.55 & 26.47 & 41.72 & 30.44 & 21.09 \\
\midrule
\multirow{4}{*}{RAS Jacobi Smoother} & Median (s) & 0.42 & 0.21 & 0.11 & 0.06 & 0.03 & 0.02 & 0.01 & 0.01 & 0.03 \\
 & 95\% CI Lower (s) & 0.42 & 0.21 & 0.11 & 0.05 & 0.03 & 0.02 & 0.01 & 0.01 & 0.03 \\
 & 95\% CI Upper (s) & 0.43 & 0.21 & 0.12 & 0.06 & 0.03 & 0.02 & 0.01 & 0.02 & 0.03 \\
 & Speedup & 1.00 & 1.99 & 3.70 & 7.25 & 15.18 & 27.88 & 43.60 & 29.60 & 15.14 \\
\bottomrule
\end{tabular}
\end{table}

\end{document}